\documentstyle[11pt]{article}

\newcommand{\BE}{\begin{equation}}
\newcommand{\EE}{\end{equation}}
\begin{document}

\begin{flushright}
Freiburg-THEP 99/09\\
August 1999 \vspace {0.8cm}\\
\end{flushright}

\begin{center}
{ \Large \bf Large rescaling of the scalar condensate, towards a Higgs-gravity
 connection ?
\footnote{Presented at the EPS-HEP99 meeting, july 1999, Tampere, Finland}}
\end{center}
\vskip 0.5cm
\begin{center}
{\large \bf J.J. van der Bij}\\
{\it Fakult\"at f\"ur Physik,}\\
{\it Albert--Ludwigs--Universit\"at Freiburg,}\\
{\it Hermann--Herder--Strasse 3,}\\ 
{\it 79104 Freiburg, Germany.}
\end{center}
\vskip 0.5cm

\begin{abstract}
In the Standard Model the Fermi constant is associated with the vacuum 
expectation value of the Higgs field $\langle\Phi\rangle$, `the condensate',
 usually believed
to be a nearly cut-off independent quantity. General arguments related to
the `triviality' of $\lambda\Phi^4$ theory in 4 space-time dimensions suggest,
however, a dramatic renormalization effect in the continuum theory.
This effect is  visible on the relatively large
lattices (such as $32^4$) available today. The result is suggestive 
of a certain `Higgs-gravity connection', as discussed
 some years ago. 
The space-time structure is determined by symmetry breaking and 
the Planck scale is essentially a rescaling
of the Fermi scale. The 
resulting picture may lead to quite substantial changes in the usual
 phenomenology associated with the Higgs particle.
\end{abstract} 
\vskip 50 pt
\par 
\par The aim of this paper is to analyze
the scale dependence of the `Higgs condensate' $\langle\Phi\rangle$, an
important subject
for any theory aiming to insert the electroweak theory
 in more ambitious 
frameworks. At the same time, it turns out that the solution of
this genuine quantum-field-theoretical problem can provide precious
insights on possible connections between the Higgs sector 
and spontaneously broken theories of gravity.
We shall restrict to the case of a 
pure $\Phi^4$ theory and comment in the end about the stability of the 
results when other interactions (such as gauge and yukawa) are present.
%
%
\par Let us define the theory in the presence of a lattice 
spacing $a \sim 1/\Lambda$ and assume
that the ultraviolet cutoff $\Lambda$ is much larger than the Higgs
mass $M_h$. Our basic problem is to relate
the bare condensate
\begin{equation}
\label{bare}
v_B(\Lambda)\equiv \langle \Phi_{\rm latt} \rangle  
\end{equation}
to the low-energy physical value $v_R$ (that in the Standard Model is
associated with the Fermi scale $v_R \sim 246$ GeV ).
 In the usual approach, one writes
\begin{equation}
\label{zeta}
         v^2_B = Z v^2_R  
\end{equation}
where $Z$ is called the `wave-function renormalization'. 
\par
As pointed out in \cite{cs}, 
in the presence of spontaneous symmetry breaking, 
where there is no smooth limit $p \rightarrow 0$, 
there are two basically different definitions of $Z$:
\par~~~~~~~~a) $Z\equiv Z_{\rm prop}$ where $Z_{\rm prop}$ is defined from
the propagator of the shifted fluctuation field 
$h(x)=\Phi(x) -\langle \Phi \rangle$, namely 
\begin{equation}
\label{prop}
       G(p^2) \sim { { Z_{\rm prop} }\over{p^2 + M^2_h} } 
\end{equation}
\par~~~~~~~~b) $Z\equiv Z_\varphi$ where $Z_\varphi$ is the rescaling 
needed in the effective potential $V_{\rm eff} (\varphi)$ 
to match the quadratic shape at its absolute minima with the Higgs mass, 
namely
\begin{equation}
\label{match}
   { { d^2V_ {\rm eff} } \over { d {\varphi}^2_R }} |_{ {\varphi}_R=v_R}=
M^2_h
\end{equation}
\par Now, by assuming `triviality' as an exact property of $\Phi^4$ theories in
four space-time dimensions 
\cite{book}, it is
well known that the continuum limit leads to $Z_{\rm prop} \to 1$. 
However, is this result relevant to evaluate the scale dependence of the 
Higgs {\it condensate} ? When dealing with
the space-time constant part of the
scalar field, it is $Z\equiv Z_\varphi$, as defined from Eq.(4), that represents
the relevant definition to be used in Eq.(2). 
\par
Now, as pointed out in 
\cite{cs}, by restricting to
those approximations to the effective potential
(say $V_{\rm eff}\equiv V_{\rm triv}$ ) that are consistent with `triviality',
since the field $h(x)$ is governed by a 
quadratic hamiltonian and $Z_{\rm prop}=1$, 
one finds a non-trivial $Z_\varphi$. Indeed, 
$V_{\rm triv}$,  given by the sum of a classical
potential and the zero-point energy of the shifted fluctuation field, is an 
extremely {\it flat} function of $\varphi_B$ implying a divergent 
$Z_\varphi\sim \ln {{\Lambda}\over{M_h}} $ in the limit 
$\Lambda \to \infty$. Thus, when properly understood,
`triviality' requires at the 
same time $Z_{\rm prop} \to 1$ and $Z_\varphi \to \infty$, implying that, in 
a true continuum theory, $v_B/v_R$ can become arbitrarily large. 
\par The existence of a non-trivial
$Z_\varphi$ for the Higgs condensate, quite distinct from the $Z_{\rm prop}$, 
associated with the finite-momentum fluctuation field, is a definite prediction
that can be tested with a precise
lattice computation. To this end a lattice simulation of the Ising limit of 
$\Phi^4$ theory was used \cite{old} to measure i)
 the zero-momentum susceptibility:
\begin{equation}
\label{chi}
{\chi}^{-1} = 
{ { d^2V_ {\rm eff} } \over {d {\varphi}^2_B }}|_{ {\varphi}_B=v_B}\equiv
{{M^2_h}\over{Z_\varphi}}
\end{equation}
and ii) the propagator of the shifted field (at Euclidean momenta $p\neq 0$).
The latter data can be fitted to the form in Eq.(3) to obtain 
$M_h$ and $Z_{\rm prop}$. 
\par The lattice data show that $Z_{\rm prop}$ is slightly less than one, and
tends to unity as the continuum limit is approached, consistently with the 
non-interacting nature of the field $h(x)$. However, the $Z_\varphi$ extracted
from the susceptibility in the broken phase
is clearly different. It shows a rapid increase above
unity and the trend is consistent with it diverging in the continuum limit. 
Absolutely no sign of such a discrepancy is found in the symmetric phase, as
expected. The first indications obtained in \cite{old} are now confirmed by
a more complete analysis \cite{cnew} performed with more statistics on the
largest lattices employed so far, such as $32^4$. The results accord well 
with the "two-Z" picture where one predicts a $Z_{\rm prop}$ slowly approaching
one and
a zero-momentum rescaling 
$Z_\varphi=M^2_h \chi$ becoming higher and higher the closer we get to the 
continuum limit.
%
%
\par
The above result, giving a large difference in the rescalings of the vacuum 
and the fluctuation fields, has been established for the model with
a single scalar field. The results have been described on the lattice.
In order to make contact with experiment, one must extend the results 
to the O(4) sigma model, which describes the Higgs sector of the standard 
model. This can be done straightforwardly when one has a description of the
effect directly in the continuum. To describe the different rescalings in the
continuum one should consider the effective Lagrangian, which is in general a
combination of many operators. We are looking for an operator that would give
rise to a rescaling of the Higgs propagator in the broken phase but not
in the unbroken phase. At the same time it should leave the constant fields
untouched and be invariant under the symmetry of the theory. This leads us 
to consider the following operator:  
$(\partial_{\mu}
\vert \Phi \vert ^2)(\partial^{\mu} \vert \Phi \vert ^2)$.
Because of the derivatives this operator leaves the constant fields untouched,
but leads to a wave function renormalization of the Higgs field after
spontaneous symmetry breaking. To make a connection with the standard model one
now simply takes $\Phi$ to be the four component Higgs field.
\par
If one adds this term to the standard model as an effective interaction, 
with a very large coefficient, the result after symmetry breaking is very
simple. In the unitary gauge one finds simply the standard model, but with 
a Higgs coupling to matter reduced by the wave function renormalization.
Ultimately, when one takes the wave function renormalization to $\infty$,
the Higgs particle does not couple to matter anymore. One therefore ends up
with the standard model without Higgs-particle. Given the precision
of present data in the weak interactions one has to ask oneself whether
this picture is in agreement with experiment. At the fundamental level there 
is clearly a problem reconciling perturbative calculations of electroweak
parameters like $\delta \rho$ with the demands of triviality in the scalar
sector. To do perturbative calculations one simply needs the Higgs matter and
self-couplings to get finite results. However presently we are only sensitive
to one-loop corrections, where these couplings do not play a role. At the 
one loop level the Higgs particle plays essentially a role as a cut-off to
the momentum integrals. Corrections to electroweak quantities behave like
$log(m_H/m_W) + c$. The logarithmic corrections are universal, in the 
sense that essentially any form of dynamical symmetry breaking would give
rise to logarithmic divergences with the same coefficients as the $log(m_H)$
terms. It is only the constants, that are unique for the normal perturbative
interpretation of the standard model. While present data indicate a low
value of the Higgs mass within the standard perturbative calculations,
they are not precise enough to determine the constant terms uniquely.
Therefore a model without a Higgs, but with a cut-off around the weak scale,
implying new interactions, cannot be excluded at present.
\par
The origin of the triviality, as found above from the lattice data, lies
in the appearance of a very large, essentially infinite, ratio of
wave function renormalizations. This naturally brings up the question
whether there might be a connection with other large ratios of parameters
that appear in fundamental physics. Here we suggest a connection with
the ratio of Planck mass and Fermi scale.
There are a number of reasons that suggest that a connection
between gravity and the Higgs sector is possible. There is for instance the 
question of the cosmological constant, which is generated by the Higgs 
potential.
Another argument is the fact, that both gravity and the Higgs particle have
a universal form of coupling to matter.  Gravity couples universally to the 
energy-momentum tensor, the Higgs particle to mass, which corresponds to
the trace of the energy-momentum tensor. However the way the Planck scale
appears is quite different from the way the weak scale appears. This
discrepancy can be resolved by assuming that also the Planck scale
arises after spontaneous symmetry breaking.

We start therefore with the Lagrangian:

$$ {\cal L} = \sqrt{g}\bigl ( \xi \Phi^{+} \Phi R -\frac {1}{2} g^{\mu \nu}
  (D_{\mu} \Phi)^{+}(D_{\nu}\Phi) -V(\Phi^{+} \Phi) - \frac {1}{4}
F_{\mu \nu} F^{\mu \nu} \bigr )\eqno(5)$$

This is the spontaneous symmetry breaking theory of gravity \cite{adler}, 
with the standard model Higgs v.e.v. as the origin of the Planck mass. 
The parameter $\xi$ is given by $\xi=1/16 \pi G_N v^2 $.
This parameter is of $O(10^{34})$ and gives rise to a large mixing between
the Higgs boson and the graviton. 
 This model was recently discussed in \cite{bij}. 
The physical content of the model becomes clear after the Weyl rescaling
 $
g_{\mu \nu} \rightarrow \frac{\kappa^2}{\xi \vert \Phi \vert ^2}
g_{\mu \nu}$, giving the Lagrangian :
$$ {\cal L} = \sqrt{g}\bigl ( \kappa^2 R -\frac{3}{2}\frac {\xi v^2}
{\vert \Phi \vert^4} 
(\partial_{\mu}
\vert \Phi \vert ^2)(\partial^{\mu} \vert \Phi \vert ^2)
-\frac {1}{2} \frac {v^2}{\vert \Phi \vert^2} (D_{\mu} \Phi^{+})
(D^{\mu} \Phi) - \frac {v^4}{\vert \Phi \vert^4}  V(\vert \Phi \vert^2)
\bigr )
\eqno(6) $$

As $\xi$ is very large indeed, this final Lagrangian is  of the right form
to be consistent with the triviality of the Higgs sector. After making the
wavefunction renormalization by the factor $(1 + 12 \xi)^{1/2}$ the coupling
of the Higgs particle to ordinary matter becomes of gravitational strength.
Also the mass of the Higgs particle becomes of the order $v^2/m_P$.
This would give rise to a Yukawa potential of gravitational strength with
a range that could be in the millimeter range. Such a force can be looked
for at the next generation of fifth force experiments.

If indeed such a close connection between gravity and the Higgs sector 
is present in nature, quite dramatic effects should be expected at the
electroweak scale, as quantum gravity would already appear at this scale.
What these effects are precisely, cannot be predicted confidently at the
present time. Typical suggestions are higher dimensions, string Regge
trajectories, etcetera. In any case it is very suggestive that the picture
of triviality in the Higgs sector, can be so naturally  described in the
spontaneous symmetry breaking theory of gravity.\\

{\bf Acknowledgements} I wish to thank Prof. M. Consoli for many illuminating
discussions on the nature of triviality. I thank the INFN, Catania for
hospitality. This work was supported by the Deutsche Forschungsgemeinschaft.\\

\end{document}